\documentstyle{black}
\input{psfig}

\def\sec{\,{\rm sec}}
\def\Gyr{\,{\rm Gyr}}

\def\rcm{\,{\rm cm}}

\def\Mpc{\,{\rm Mpc}}

\def\cmm2{{\,\rm cm^{-2}}}
\def\cm2{{\,{\rm cm}^2}}
\def\cmm3{{\,{\rm cm}^{-3}}}
\def\gcmm3{{\,{\rm g\,cm^{-3}}}}
\def\kms{\,{\rm km\,s^{-1}}}

\def\la{\mathrel{\mathpalette\fun <}}
\def\ga{\mathrel{\mathpalette\fun >}}
\def\fun#1#2{\lower3.6pt\vbox{\baselineskip0pt\lineskip.9pt
  \ialign{$\mathsurround=0pt#1\hfil##\hfil$\crcr#2\crcr\sim\crcr}}}

\begin{document}

\title{COSMOLOGY SOLVED?  MAYBE.}

\author{
Michael S. TURNER\\
{\it Departments of Astronomy \& Astrophysics and of Physics \\
Enrico Fermi Institute, The University of Chicago \\
Chicago, IL~~60637-1433, USA}\\
\\
{\it NASA/Fermilab Astrophysics Center\\
Fermi National Accelerator Laboratory\\
Batavia, IL~~60510-0500, USA\\
mturner@oddjob.uchicago.edu}
}

\maketitle

\section*{Abstract}
For two decades the hot big-bang model as been referred to as the
standard cosmology -- and for good reason.  For just as long
cosmologists have known that there are fundamental questions
that are not answered by the standard cosmology and point to
a grander theory.  The best candidate for that grander
theory is inflation + cold dark matter; it can extend
our understanding of the Universe back to $10^{-32}\sec$.
There is now prima facie evidence that supports the
two basic tenets of this new paradigm:  flat Universe and
scale-invariant spectrum of Gaussian density perturbations.  An avalanche of
high-quality cosmological observations will soon make this case
stronger or will break it.  If inflation + cold dark matter
is correct, then there are new, fundamental questions
to be answered, most notably the nature of the
dark energy that seems to account for 60\% of the
critical density and how inflation fits into a unified
theory of the forces and particles.
These are exciting times in cosmology!

\section{1998, A Memorable Year for Cosmology}

The birth of the hot big-bang model dates back to the work of Gamow
and his collaborators in the 1940s.  The emergence of the hot big-bang
as the standard cosmology began in the late 1960s, with the discovery of
the microwave background radiation, the establishment of its black-body
spectrum, and the success of big-bang nucleosynthesis.  By the 1970s,
the hot big-bang was being referred to as the standard
cosmology.  Today, it is well established and provides an accounting
of the Universe from a fraction of a second after the beginning
when the Universe was a hot, smooth soup of quarks and leptons to the
present, some $13\Gyr$ later.  The standard cosmology rests upon
three strong observational pillars:  the expansion of the Universe; the cosmic
microwave background radiation (CBR); and the abundance pattern
of the light elements, D, $^3$He, $^4$He, and $^7$Li,
produced seconds after the bang (see e.g., Peebles et al, 1991).

The standard cosmology leaves fundamental questions
unexplained:  the matter/antimatter asymmetry, the origin of the smoothness
and flatness of the Universe, the nature and origin of the
primeval density inhomogeneities that seeded all the structure
in the Universe, the quantity and composition of the dark matter
that holds the Universe together, and the nature of the big-bang
event itself.
This has motivated the search for a more expansive cosmological theory.

In the 1980s, a new paradigm emerged, deeply rooted in fundamental
physics with the potential to extend our understanding of the Universe
back to $10^{-32}\sec$ and to address the fundamental questions
poised by the hot big-bang model.  That paradigm, known as inflation
+ cold dark matter, holds that most of the dark matter consists of
slowly moving elementary particles (cold dark matter),
that the Universe is flat and
that the density perturbations that seeded all the structure seen
today arose from quantum mechanical fluctuations on scales of $10^{-23}
\rcm$ or smaller.  It took awhile for the observers and experimentalists
to take this theory seriously enough to try to disprove it, and in the 1990s
it began to be tested in a serious way.

\begin{figure}
\centerline{\psfig{figure=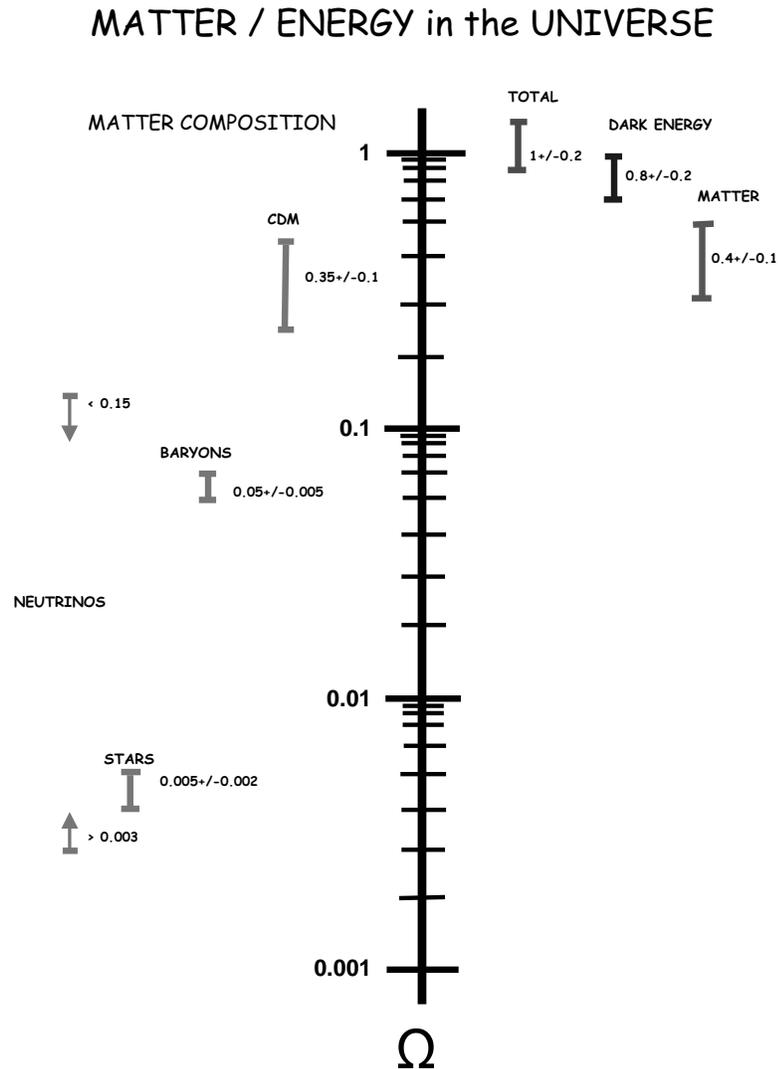,width=4in}}
\caption{Summary of matter/energy in the Universe.
The right side refers to an overall accounting of matter
and energy; the left refers to the composition of the matter
component.  The upper limit to mass density contributed
by neutrinos is based upon the failure of the hot dark
matter model and the lower limit follows from the
evidence for neutrino oscillations (Fukuda et al, 1998).
}
\label{fig:omega_sum}
\end{figure}

1998 could prove to be a watershed year in cosmology, as important
as 1964, when the CBR was discovered.  The crucial
new data include a precision measurement of the
density of ordinary matter and of the total amount
of matter, both derived from a measurement of
the primeval deuterium abundance and the theory of BBN; and
the first fine-scale (down to $0.3^\circ$) measurements of
the anisotropy of the CBR; and a measurement of the deceleration
of the Universe based upon distance measurements of
type Ia supernovae (SNe1a) out to redshift of close to unity.
Together, these measurements, which are
harbingers for the precision era of cosmology that is coming,
provide the first plausible, complete accounting of the matter/energy
density in the Universe and evidence that the primeval density perturbations
arose from quantum fluctuations during inflation.
In addition, there exists a body of
evidence in support of the cold dark matter theory of structure
formation.

The accounting of matter and energy goes like this
(in units of the critical density):  light neutrinos,
at least 0.3\%; bright stars and related
material, 0.5\%; baryons, 5\%; cold dark matter, 35\%; and vacuum
energy (or something similar), 60\%; for a total equalling
the critical density (see Fig.~1).
The recently measured primeval
deuterium abundance (Burles \& Tytler, 1998)
and the theory of big-bang nucleosynthesis
accurately determine the baryon density (Schramm \& Turner, 1998),
$\Omega_B = (0.02\pm 0.002)
h^{-2} \simeq 0.05$ (for $h=0.65$).  Using the cluster baryon
fraction, determined from x-ray measurements, $f_B = M_{\rm baryon}/
M_{\rm TOT} = 0.07\pm 0.007$ (Evrard, 1996), and assuming that clusters provide
a fair sample of matter in the Universe, $\Omega_B/\Omega_M
= f_B$, it follows that $\Omega_M = (0.3\pm 0.05)h^{-1/2} \simeq
0.4\pm 0.1$.  That $\Omega_M \gg \Omega_B$ is strong evidence
for nonbaryonic dark matter; the leading candidates are axions,
neutralinos and neutrinos.

The position of the first acoustic peak in the angular power spectrum
of temperature fluctuations of the CBR
is a sensitive indicator of the curvature of the Universe:
$l_{\rm peak} \simeq 200/\sqrt{\Omega_0}$, where $R_{\rm curv}^2
= H_0^{-2}/|\Omega_0 -1|$.  Measurements now span
multipole number $l=2$ to around $l=1000$ (see Fig.~2);
while the data do not yet speak definitively, it is clear that $\Omega_0 \sim
1$ is preferred.  Several experiments have new results
around $l=30 -300$, and should be reporting them soon.  Ultimately, the MAP
(launch in 2000) and Planck (launch in 2007)
satellites will cover $l=2$ to $l=3000$ with precision
limited essentially by sampling variance, and should determine
$\Omega_0$ to a precision of 1\% or better.

The same angular power spectrum that indicates $\Omega_0\sim 1$
also provides evidence that the primeval density perturbations
are of the kind predicted by inflation.  The inflation-produced
Gaussian curvature fluctuations lead to an angular
power spectrum with a series of well defined acoustic peaks.  While the
data at best define the first peak, they are good enough to
exclude models where the density perturbations are isocurvature
(e.g., cosmic strings and textures):  in these models the predicted
spectrum is devoid of acoustic peaks (Allen et al, 1997; Pen et al,
1997).

The oldest approach to determining $\Omega_0$ is by measuring
the deceleration of the expansion.  Sandage's deceleration parameter,
$q_0 \equiv -({\ddot R}/R)/H_0^2 = {\Omega_0\over 2}[1+3p/\rho ]$,
depends upon both $\Omega_0$ and the equation of state.  Accurate
measurements of the (luminosity) distance as a function of
redshift allow the deceleration to be determined.  Accurate
distant measurements to some fifty or so SNe1a, with redshifts as
large as one, carried out by two groups (Riess et al, 1998;
Perlmutter et al, 1998) indicate that the Universe is speeding
up, not slowing down (i.e., $q_0<0$).  The simplest explanation is
a cosmological constant, with $\Omega_\Lambda \sim 0.6$.  This
result fits neatly with the CBR determination that $\Omega_0 =1$
and dynamical measures that indicate $\Omega_M\sim 0.4$:  the
``missing energy'' exists in a smooth component that cannot
clump and thus is not found in clusters of galaxies.

While the evidence for inflation + cold dark matter
is not definitive and we should be cautious,
1998 could well mark a turning point in cosmology as important
as 1964.  Recall, after the discovery of the CBR
it took a decade or more to firmly
establish the cosmological origin of the CBR and
the hot big-bang cosmology as the standard cosmology.

\section{Inflation + Cold Dark Matter}

\begin{figure}
\centerline{\psfig{figure=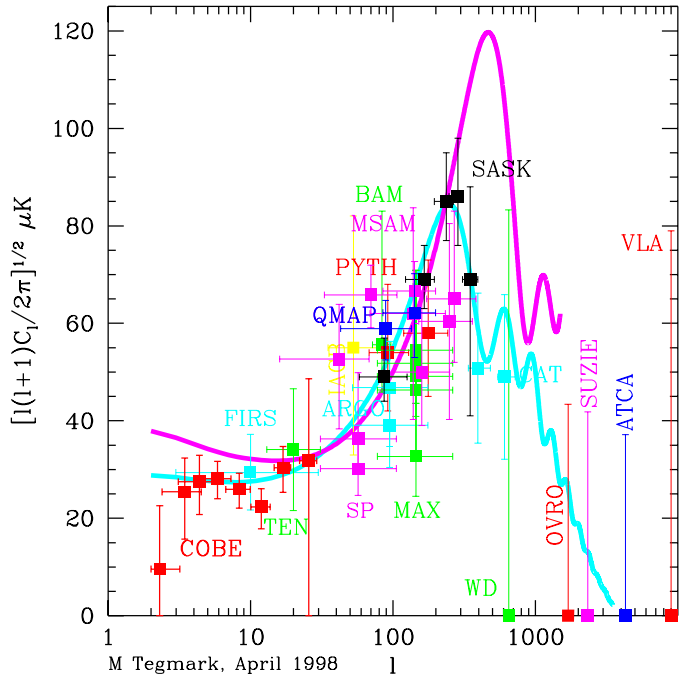,width=3.5in}}
\caption{Summary of current CBR anisotropy measurements, where
the temperature variation across the sky has been expanded
in spherical harmonics, $\delta T(\theta , \phi ) = \sum_i a_{lm}Y_{lm}$
and $C_l \equiv \langle |a_{lm}|^2\rangle$.  The curves
illustrate CDM models with $\Omega_0 = 1$ and $\Omega_0 =0.3$.
Note the preference of the data for a flat Universe
(Figure courtesy of M. Tegmark).}
\label{fig:cbr_today}
\end{figure}

Inflation has revolutionized the way cosmologists view the Universe
and provides the current working hypothesis for extending the
standard cosmology.  It explains how a region of size much, much
greater than our Hubble volume could have become smooth and flat
without recourse to special initial conditions (Guth 1981), as well as
the origin of the density inhomogeneities
needed to seed structure (Hawking, 1982; Starobinsky, 1982;
Guth \& Pi, 1982; and Bardeen et al, 1983).
Inflation is based upon
well defined, albeit speculative physics -- the semi-classical evolution
of a weakly coupled scalar field -- and this physics may well
be connected to the unification of the particles and forces of Nature.

It would be nice if there were a standard model of inflation, but
there isn't.   What is important, is that almost all inflationary
models make three very testable predictions:  flat
Universe, nearly scale-invariant
spectrum of Gaussian density perturbations, and nearly scale-invariant
spectrum of gravitational waves.  These three predictions allow the
inflationary paradigm to be decisively tested.  While the gravitational
waves are an extremely important and challenging test, I do not have
space to mention them again here (see e.g., Turner 1997).

The tremendous expansion that occurs during inflation
is key to its beneficial effects and robust predictions:
A small, subhorizon-sized bit
of the Universe can grow large enough to encompass the entire
observable Universe and much more.  Because all that
we can see today was once so extraordinarily small, it appears flat
and smooth.  This is unaffected
by the expansion since then and so the Hubble radius today is
much, much smaller than the curvature radius, implying $\Omega_0
=1$.  Lastly, the tremendous expansion stretches quantum fluctuations
on truly microscopic scales ($\la 10^{-23}\rcm$)
to astrophysical scales ($\ga \Mpc$).

The curvature perturbations created by inflation
are characterized by two important features: 1) they are
almost scale-invariant, which refers to the fluctuations in
the gravitational potential being independent
of scale -- and not the density perturbations themselves;
2) because they arise from fluctuations in an essentially noninteracting
quantum field, their statistical properties are that of
a Gaussian random field.

Scale invariance specifies the dependence of the spectrum of density
perturbations upon scale.  The normalization (overall amplitude) depends
upon the specific inflationary model (i.e., scalar-field potential).
Ignoring numerical factors for the moment, the fluctuation amplitude
is given by:  $\delta \phi\simeq (\delta \rho /\rho )_{\rm HOR} \sim
V^{3/2}/m_{\rm PL}^3 V^\prime$.
(The amplitude of the density perturbation on a given scale
at horizon crossing is equal to the fluctuation in the gravitational
potential $\delta \phi$.)  To be consistent with the COBE
measurement of CBR anisotropy
on the $10^\circ$ scale, $\delta \phi$ must be around $2\times 10^{-5}$.
Not only did COBE produce the first evidence for the existence of
the density perturbations that seeded all structure
(Smoot et al, 1992), but also,
for a theory like inflation that predicts
the shape of the spectrum of density perturbations,
it provides the overall normalization that
fixes the amplitude of density perturbations on all scales.
The COBE normalization began precision testing of inflation.

\section{Inflation + CDM in the Era of Precision Cosmology}

As we look forward to the abundance (avalanche!) of high-quality observations
that will test Inflation + CDM, we have to make sure the predictions
of the theory match the precision of the data.  In
so doing, CDM + Inflation becomes a ten (or more) parameter
theory.  For astrophysicists, and especially cosmologists,
this is daunting, as it may seem that a ten-parameter
theory can be made to fit any set of observations.  This is
not the case when one has the quality and quantity of data
that will be coming.  The standard model of particle physics offers
an excellent example:  it is
a nineteen-parameter theory and because of the high-quality of
data from experiments at Fermilab's Tevatron, SLAC's SLC,
CERN's LEP and other facilities it has been rigorously tested
and the parameters measured to a precision of better than 1\%
in some cases.  My worry as an inflationist is not that many different
sets of parameters will fit the upcoming data, but rather that
no set of parameters will!

In fact, the ten parameters of CDM + Inflation
are an opportunity rather than a curse:  Because the parameters
depend upon the underlying inflationary model and fundamental
aspects of the Universe, we have the very real possibility of learning
much about the Universe and inflation.  The ten parameters
can be organized into two groups:  cosmological
and dark-matter (Dodelson et al, 1996).

\smallskip
\centerline{\it Cosmological Parameters}
\begin{enumerate}

\item $h$, the Hubble constant in units of $100\kms\Mpc^{-1}$.

\item $\Omega_Bh^2$, the baryon density.  Primeval deuterium
measurements and together with the theory of BBN imply:
$\Omega_Bh^2 = 0.02 \pm 0.002$.

\item $n$, the power-law index of the scalar density perturbations.
CBR measurements indicate $n=1.1\pm 0.2$; $n=1$ corresponds to
scale-invariant density perturbations.  Several popular
inflationary models predict $n\simeq 0.95$; range of predictions
runs from $0.7$ to $1.2$ (Lyth \& Riotto, 1996).

\item $dn/d\ln k$, ``running'' of the scalar index with comoving scale
($k=$ wavenumber).  Inflationary models predict a value of
${\cal O}(\pm 10^{-3})$ or smaller (Kosowsky \& Turner, 1995).

\item $S$, the overall amplitude squared of density perturbations,
quantified by their contribution to the variance of the
CBR quadrupole anisotropy.

\item $T$, the overall amplitude squared of gravity waves,
quantified by their contribution to the variance of the
CBR quadrupole anisotropy.  Note, the COBE normalization determines
$T+S$ (see below).

\item $n_T$, the power-law index of the gravity wave spectrum.
Scale-invariance corresponds to $n_T=0$; for inflation, $n_T$
is given by $-{1\over 7}{T\over S}$.

\end{enumerate}

\smallskip
\centerline{\it Dark-matter Parameters}

\begin{enumerate}

\item $\Omega_\nu$, the fraction of critical density in neutrinos
($=\sum_i m_{\nu_i}/90h^2$).  While the hot dark matter theory of structure
formation is not viable, it is possible that a small fraction of
the matter density exists in the form of neutrinos.
Further, small -- but nonzero -- neutrino masses are
a generic prediction of theories that unify the
strong, weak and electromagnetic interactions  -- and the
Super-Kamiokande Collaboration has presented
evidence that the at least one of the neutrino species has a
mass of greater than about 0.1\,eV, based upon the deficit
of atmospheric muon neutrinos (Fukuda et al, 1998).

\item $\Omega_X$, the fraction of critical density in a smooth component
of unknown composition and negative pressure ($w_X \la -0.3$).  There is
mounting evidence for such a component, with the simplest example being
a cosmological constant ($w_X = -1$).

\item $g_*$, the quantity that counts the number of ultra-relativistic
degrees of freedom (around the time of matter-radiation
equality).  The standard cosmology/standard
model of particle physics predicts $g_* = 3.3626$ (photons in the
CBR + 3 massless neutrino species with temperature $(4/11)^{1/3}$
times that of the photons).  The amount of radiation controls when
the Universe became matter dominated and thus affects the present
spectrum of density inhomogeneity.

\end{enumerate}

\subsection{Present status of Inflation + CDM}

A useful way to organize the different CDM models is by their
dark-matter content; within each CDM family, the cosmological
parameters vary.  One list of models is:

\begin{enumerate}

\item sCDM (for simple):  Only CDM and baryons; no additional
radiation ($g_*=3.36$).  The original standard CDM is a member
of this family ($h=0.50$, $n=1.00$, $\Omega_B=0.05$), but is
now ruled out (see Fig.~3).

\item $\tau$CDM:  This model has
extra radiation, e.g., produced by the decay of an unstable
massive tau neutrino (hence the name); here we take $g_* = 7.45$.

\item $\nu$CDM (for neutrinos):  This model has a dash of hot
dark matter; here we take $\Omega_\nu = 0.2$ (about 5\,eV
worth of neutrinos).

\item $\Lambda$CDM (for cosmological constant):  This model has
a smooth component in the form of a cosmological constant; here
we take $\Omega_\Lambda = 0.6$.

\end{enumerate}

\begin{figure}
\centerline{\psfig{figure=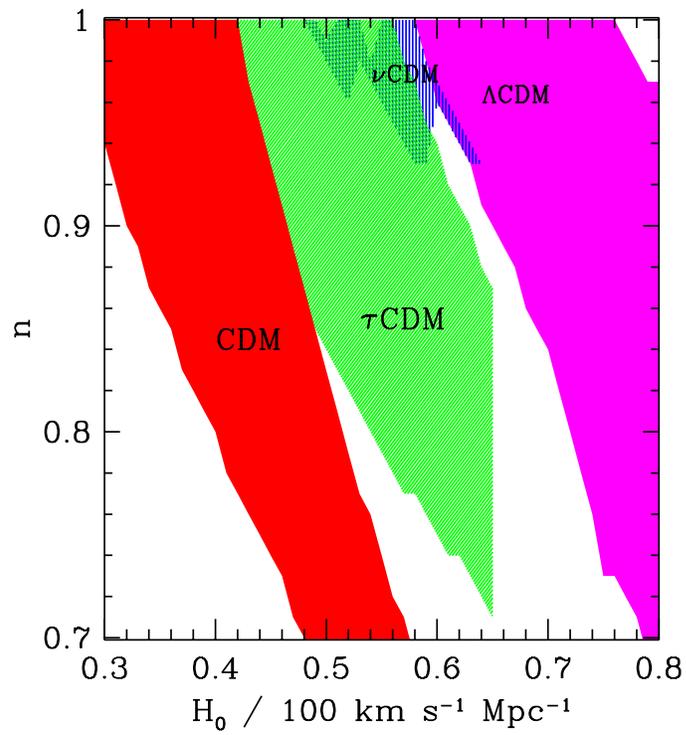,width=3.5in}}
\caption{Summary of viable CDM models, based upon
CBR anisotropy and determinations of the present
power spectrum of inhomogeneity (Dodelson et al, 1996).}
\label{fig:cdm_sum}
\end{figure}

Figure 3 summarizes the viability of these different CDM models,
based upon CBR measurements and current determinations of
the present power spectrum of inhomogeneity derived from
redshift surveys.   sCDM is only viable for low values of the
Hubble constant (less than $55\kms\Mpc^{-1}$) and/or
significant tilt (deviation from scale invariance); the region
of viability for $\tau$CDM is similar to sCDM, but shifted
to larger values of the Hubble constant (as large as
$65\kms\Mpc^{-1}$).  $\nu$CDM has an island of viability
around $H_0\sim 60\kms\Mpc^{-1}$ and $n\sim 0.95$.  $\Lambda$CDM
can tolerate the largest values of the Hubble constant.

Considering other relevant data too -- e.g.,
age of the Universe, determinations of $\Omega_M$,
measurements of the Hubble constant, and limits to
$\Omega_\Lambda$ -- $\Lambda$CDM emerges as the
`best-fit CDM model' (Krauss \& Turner, 1995;
Ostriker \& Steinhardt, 1995; Liddle et al, 1996);
see Fig.~4.
Moreover, its `smoking gun signature,'
negative $q_0$, has apparently been confirmed (Riess et al, 1998;
Perlmutter et al, 1998).  Despite my general
enthusiasm, I would caution that it is premature
to conclude that $\Lambda$CDM is anything but the model
to take aim at.

\begin{figure}
\centerline{\psfig{figure=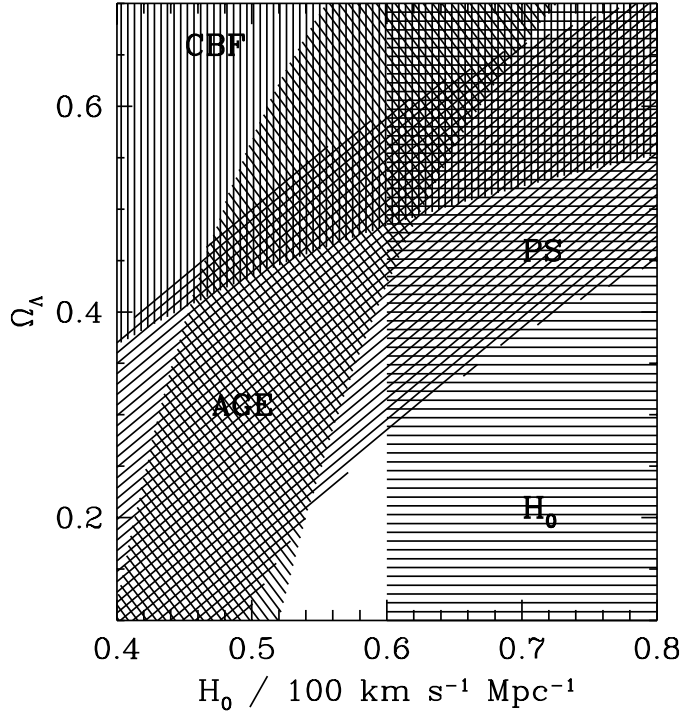,width=3.5in}}
\caption{Constraints used to determine the best-fit CDM model:
PS = large-scale structure + CBR anisotropy; AGE = age of the
Universe; CBF = cluster-baryon fraction; and $H_0$= Hubble
constant measurements.  The best-fit model, indicated by
the darkest region, has $h\simeq 0.60-0.65$ and $\Omega_\Lambda
\simeq 0.55 - 0.65$.}
\label{fig:best_fit}
\end{figure}

\section{Checklist for the Next Decade}

As I have been careful to stress the basic tenets of Inflation
+ Cold Dark Matter have not yet been confirmed definitively.
However, a flood of high-quality cosmological data is
coming, and could make the case in the next decade.
Here is my version of how ``maybe'' becomes ``yes.''

\begin{itemize}

\item Map of the Universe at 300,000 yrs.  COBE mapped
the CMB with an angular resolution of around $10^\circ$;
two new satellite missions, NASA's MAP (launch 2000)
and ESA's Planck Surveyor (launch 2007), will map the
CMB with 100 times better resolution ($0.1^\circ$). From
these maps of the Universe as it existed at a
simpler time, long before the first stars and
galaxies, will come a gold mine of information:
Among other things, a definitive measurement of $\Omega_0$;
a determination of the Hubble constant to a precision of
better than 5\%;
a characterization of the primeval lumpiness; and possible
detection of the relic gravity waves from inflation.
The precision maps of the CMB that will be made are crucial
to establishing Inflation + Cold Dark Matter.

\item Map of the Universe today.  Our knowledge of the structure
of the Universe is based upon maps constructed from the positions
of some 30,000 galaxies in our own backyard.  The Sloan Digital
Sky Survey will
produce a map of a representative portion of the Universe,
based upon the positions of a million galaxies.
The Anglo-Australian 2-degree Field survey will determine the position of
several hundred thousand galaxies.  These surveys will
define precisely the large-scale structure that exists today,
answering questions such as, ``What are the largest structures
that exist?''  Used together with
the CMB maps, this will definitively test the Cold
Dark Matter theory of structure formation, and much more.

\item Present expansion rate $H_0$.  Direct measurements
of the expansion rate using standard candles, gravitational time
delay, SZ imaging and the CMB maps will pin down the elusive
Hubble constant once and for all.  It is the fundamental parameter
that sets the size -- in time and space -- of the observable Universe.
Its value is critical to testing the self consistency of Cold
Dark Matter.

\item Cold dark matter.  A key element of theory is the
cold dark matter particles that hold the Universe together;
until we actually
detect cold dark matter particles, it will be difficult to argue
that cosmology is solved.
Experiments designed to detect the dark matter that
holds are own galaxy together are now operating with sufficient
sensitivity to detect both neutralinos and axions.
In addition, experiments at particle accelerators (Fermilab
and CERN) will be hunting for the neutralino and its other
supersymmetric cousins.

\item Nature of the dark energy.  If the Universe is indeed accelerating,
then most of the critical density exists in the form of dark energy.  This
component is poorly understood.  Vacuum energy is only the
simplest possibly for the smooth dark component; there are
other possibilities:  frustrated topological defects
or an evolving scalar field (see e.g., Caldwell et al, 1998;
Turner \& White, 1997).   Independent evidence for the existence
of this dark energy, e.g., by CMB anisotropy, the SDSS and 2dF
surveys, or gravitational
lensing, is crucial for verifying the accounting of matter and energy
in the Universe I have advocated.  Additional measurements of SNe1a could
help shed light on the precise nature of the dark energy.  The dark
energy problem is not only of great importance for cosmology, but
for fundamental physics as well.  Whether it is vacuum energy or
quintessence, it is a puzzle for fundamental physics and possibly
a clue about the unification of the forces and particles.

\end{itemize}

\section{New Questions; Some Surprises?}

Will cosmologists look back on 1998 as a year that rivals
1964 in importance?  I think it is quite possible.  In
any case, the flood of data that is coming will make the
next twenty years in cosmology very exciting.  It could be
that my younger theoretical colleagues will get their wish --
inflation + cold dark matter is falsified and it's back
to the drawing board.  Or, it may be that it is roughly correct,
but the real story is richer and even more interesting.
This happened in particle
physics.  The quark model of the 1960s was based upon
an approximate $SU(3)$ global flavor symmetry,
which shed no light on the dynamics
of how quarks are held together.  The standard model of particle
physics that emerged and which provides a fundamental description of
physics at energies less than a few hundred GeV,
is based upon the $SU(3)$ color gauge theory of quarks and gluons (QCD)
and the $SU(2)\otimes U(1)$ gauge theory of the electroweak interactions.
The difference between global and local $SU(3)$ symmetry was profound.

Even if Inflation + Cold Dark Matter does pass the series of stringent
tests that will confront it in the next decade, there will
be questions to address and issues to work out.  Exactly how does
inflation work and fit into the scheme of the unification of the
forces and particles?  Does the quantum gravity era of cosmology,
which occurs before inflation, leave a detectable imprint on the
Universe?  What is the topology of the Universe and are there
additional spatial dimensions?  Precisely how did the excess of matter
over antimatter develop?  What happened before inflation?  What
does Inflation + Cold Dark Matter teach us about the unification
of the forces and particles of Nature?  We live in exciting times!

\section{References}

\vspace{1pc}

\re
1. Allen, B. et al, 1997, Phys. Rev. Lett. 79, 2624.

\re
2. Bardeen, J., P. J. Steinhardt, and M. S. Turner, 1983, Phys. Rev. D 28, 679.

\re
3. Burles, S. \& Tytler, D. 1998, Astrophys. J. 499, 699.

\re
4. Caldwell, R., Dave, R., \& Steinhardt, P.J. 1998, Phys. Rev. Lett. 80, 1582.

\re
5. Dodelson, S., E.I. Gates, and M.S. Turner, 1996, Science 274, 69.

\re
6. Evrard, A.E. 1996, MNRAS 292, 289.

\re
7. Fukuda, Y. et al (SuperKamiokande Collaboration) 1998,
Phys. Rev. Lett. 81, 1562.

\re
8. Guth, A., 1982, Phys. Rev. D 23, 347.

\re
9. Guth, A. and S.-Y. Pi, 1982, Phys. Rev. Lett. 49, 1110.

\re
10. Hawking, S.W. 1982, Phys. Lett. B 115, 295.

\re
11. Kosowsky, A. \& Turner, M.S. 1995, Phys. Rev. D 52, R1739.

\re
12. Krauss, L. \& M.S. Turner, 1995, Gen. Rel. Grav. 27, 1137.

\re
13. Liddle, A.R. et al 1996, MNRAS 282, 281.

\re
14. Lyth, D. H. \& Riotto, A. 1998, Phys. Rep., in press.

\re
15. Ostriker, J.P. \& Steinhardt, P.J. 1995, Nature 377, 600.

\re
16. Peebles, P.J.E., Schramm, D. N., Turner, E. L., \& Kron, R.G. 1991,
Nature 352, 769.

\re
17. Pen, U.-L. et al, 1997, Phys. Rev. Lett. 79, 1611.

\re
18. Perlmutter, S. et al 1998, Astrophys. J., in press.

\re
19. Reiss, A. et al 1998, Astron. J., in press.

\re
20. Schramm, D.N. \& Turner, M.S. 1998, Rev. Mod. Phys. 70, 303.

\re
21. Smoot, G. et al 1992, Astrophys. J. 396, L1.

\re
22.  Starobinskii, A.A., 1982, Phys. Lett. B 117, 175.

\re
23. Turner, M.S. 1997, Phys. Rev. 55, R435.

\re
24. Turner, M.S. \& White, M. 1997, Phys. Rev. D 56, R4439.

\vspace{1pc}

\end{document}